\documentclass[aps, prd, amsmath, floats, floatfix, twocolumn, nofootinbib, superscriptaddress, showpacs]{revtex4-1}
\usepackage{graphicx}
\usepackage{color}
\usepackage{latexsym}
\usepackage{verbatim}

\newcommand{\be}{\begin{eqnarray}}
\newcommand{\ee}{\end{eqnarray}}
\newcommand{\rar}{\rightarrow}

\begin{document}

\title{Search for astrophysical rotating Ellis wormholes with X-ray reflection spectroscopy}

\author{Menglei Zhou} 
\affiliation{Center for Field Theory and Particle Physics and Department of Physics, Fudan University, 200433 Shanghai, China}

\author{Alejandro Cardenas-Avendano} 
\affiliation{Programa de Matem\'atica, Fundaci\'on Universitaria Konrad Lorenz, 110231 Bogot\'a, Colombia}
\affiliation{Center for Field Theory and Particle Physics and Department of Physics, Fudan University, 200433 Shanghai, China}

\author{Cosimo Bambi}
\email[Corresponding author: ]{bambi@fudan.edu.cn}
\affiliation{Center for Field Theory and Particle Physics and Department of Physics, Fudan University, 200433 Shanghai, China}
\affiliation{Theoretical Astrophysics, Eberhard-Karls Universit\"at T\"ubingen, 72076 T\"ubingen, Germany}

\author{Burkhard Kleihaus}
\affiliation{Institut f\"ur Physik, Carl von Ossietzky Universit\"at Oldenburg, 26111 Oldenburg, Germany}

\author{Jutta Kunz}
\affiliation{Institut f\"ur Physik, Carl von Ossietzky Universit\"at Oldenburg, 26111 Oldenburg, Germany}

\date{\today}

\begin{abstract}
Recently, two of us have found numerically rotating Ellis wormholes as solutions of 4-dimensional Einstein gravity coupled to a phantom field. In this paper, we investigate possible observational signatures to identify similar objects in the Universe. These symmetric wormholes have a mass and are compact, so they may look like black holes. We study the iron line profile in the X-ray reflection spectrum of a thin accretion disk around rotating Ellis wormholes and we find some specific observational signatures that can be used to distinguish these objects from Kerr black holes. We simulate some observations with XIS/Suzaku assuming typical parameters for a bright AGN and we conclude that current X-ray missions cannot apply strong constraints.
\end{abstract}


\maketitle


\section{Introduction}

Lorentzian wormholes (WHs) are hypothetical topologically non-trivial structures of the spacetime connecting either two 
faraway regions of our Universe or two different universes in multiverse theories~\cite{wh1,wh2,wh3,wh4,wh5}. Unlike 
black holes (BHs) and other kinds of WHs, they have no horizon, and therefore they would permit one to travel from one 
region to another as spacetime shortcuts. There are currently no fundamental principles that can rule out the existence 
of similar structures, which are indeed allowed both in general relativity and in alternative theories of gravity under 
certain conditions (see, for instance, Refs.~\cite{wh_t1,wh_t2,wh_t3,wh_t4,wh_t5,wh_t6,wh_t7,wh_t8}). In the spirit 
that whatever is not forbidden can potentially exist, several authors have recently studied observational signatures to 
search for astrophysical WHs in the Universe~\cite{obs1,obs1a,obs1b,obs2,obs3,obs4,obs5,obs6,obs7,obs8}.

The simplest Lorentzian WH is that proposed by Ellis~\cite{wh1,wh3}, which describes a spherically symmetric and 
non-rotating WH with vanishing mass. Rotating Ellis WHs (REWHs) have been found in Ref.~\cite{kk}. They represent the 
first example of regular rotating WH solutions in 4-dimensional general relativity. These solutions were obtained from 
Einstein gravity coupled to a phantom field $\Phi$, namely a scalar field with a reversed sign in front
of its kinetic term. The action reads~\cite{kk}
\be
S = \int \left[ \frac{R}{16\pi}
+ \frac{1}{2 m_0} \left(\partial_\mu \Phi\right) 
\left(\partial^\mu \Phi\right) \right] \sqrt{-g} d^4x \, ,
\ee
where $R$ is the scalar curvature, $g$ is the determinant of the metric, and $m_0$ is a mass scale. We employ natural units 
in which $G_{\rm N} = c = \hbar = 1$ and also $m_0=1$ .

The line element for the REWHs reads~\cite{kk}
\be\label{eq:metric}
ds^{2} &=& -f dt^{2} + \frac{\nu}{f}\left(d\eta^{2}+hd\theta^{2}\right) \nonumber\\
&& +\frac{h\sin^{2}\theta}{f} \left(d\phi-\omega dt\right)^{2}  \, ,
\ee
where $f$, $\nu$, and $\omega$ are functions only of $\eta$ and $\theta$, $h=\eta^{2}+\eta_{0}^{2}$ is an auxiliary 
function and $\eta_{0}$ is a scaling factor. The coordinate $\eta$\footnote{The coordinate transformation 
between $\eta$ and the radial coordinate $r$ of the Boyer-Lindquist form of the Kerr metric can be found in 
Appendix A of Ref.~\cite{coords}.} ranges from $- \infty$ to $+ \infty$ and the limits 
$\eta \rar \pm \infty$ correspond to two distinct asymptotically flat regions. The throat of the WH is located at the 
hypersurface $\eta = 0$. The functions	 $f$, $\nu$, and $\omega$ can only be calculated numerically~\cite{kk}.

The angular velocity at the throat is
\be
\omega(\eta = 0) \equiv \omega_{0} \, ,
\ee
and relates the mass $M$ and the spin angular momentum $J$ of the WH by the formula~\cite{kk}
\be
M = 2 \omega_{0} J \, .
\ee 

As in Ref.~\cite{kk}, here we only focus on symmetric WHs, where $f$ and $\nu$ are even functions of $\eta$, 
i.e. solutions with spacelike sections which are symmetric under the interchange of the asymptotically flat regions. For 
our discussion, it is convenient to introduce the dimensionless rotational velocity of the throat $v_{e}$ 
\be
v_{e}=\frac{R\omega_{0}}{c} \, ,
\ee
where $R$ is the equatorial radius of the throat
\be
R=\left.\sqrt{g_{\varphi\varphi}}\right|_{\eta=0,\theta=\pi/2} \, . 
\ee

The aim of this paper is to study the iron line profile in the X-ray reflection spectrum of a thin accretion disk 
around REWHs and to find some specific observational signatures that can be used to distinguish 
these objects from Kerr BHs. This technique has been applied to different scenarios and, compared with other 
approaches, can also provide strong constraints, see for instance, Refs.~\cite{v2-1,v2-2,acajc2,eclipse2}.

This paper is organized as follows. Section~\ref{sec:AWH} reviews the basic aspects of astrophysical WHs. In Section~\ref{sec:Simu}, we present our simulations. Section~\ref{sec:Discu} is devoted to the discussion of our results. In Section~\ref{sec:SumaConc}, there are summary and conclusions.

\section{Astrophysical Wormholes} \label{sec:AWH}

Non-rotating Ellis WHs have vanishing mass, but they can affect the trajectories of particles. The search for these objects in the Universe has thus focused on the study of their gravitational lensing properties~\cite{obs1,obs2,obs3,obs4,obs6}.

REWHs have instead a mass and, being compact, they may look like BHs. The existence of bound orbits permits them to have an accretion disk and us to consider the properties of the electromagnetic radiation emitted by the gas in the disk. The continuum-fitting and the iron line methods are currently the two main techniques to test the spacetime geometry around BH candidates~\cite{rev1,rev2}. If astrophysical BH candidates, or at least some of them, were instead REWHs, the study of the thermal or of the reflection spectrum of these objects may reveal their actual nature.

Since the thermal spectrum is often not very informative about the spacetime geometry around a BH candidate~\cite{kong} and is restricted to stellar-mass objects\footnote{The temperature of a thin accretion disk around a compact object roughly scales as $M^{-0.25}$. For $M = 10$~$M_\odot$, the spectrum is in the soft X-ray band and the continuum-fitting method can be applied. For objects of millions or billions of $M_\odot$, the spectrum falls in the UV/optical bands and extinction and dust absorption
limit the ability to make an accurate measurement.}, here we study possible observational signatures of REWHs in the 
iron line profile. In the fits of real data, one has to consider the whole reflection spectrum, but eventually it is the 
iron line that provides information about the metric around the compact object and therefore it makes sense to focus on 
the iron line only in an exploratory work like this one.

The reflection component in the X-ray spectrum of BH candidates is originated by the illumination of a cold accretion disk by a hot corona~\cite{fe1,fe2}. The accretion disk is described by the Novikov-Thorne model~\cite{ntm,ntm2}: the disk is in the equatorial plane, the particles of the gas follow nearly geodesic circular orbits, and the inner edge of the disk is at the radius of the innermost stable circular orbit (ISCO). In the case of REWHs, the ISCO is at the WH throat for corotating orbits~\cite{kk}.

The iron line around WHs has been already studied in Ref.~\cite{obs5}, but in that case it was considered an {\it ad 
hoc} WH 
metric. Here we have a theoretically motivated scenario, as we consider exact solutions of Einstein gravity coupled to a 
phantom field. As we will see in the next section, some REWHs share the features of the Lorentzian WHs studied 
in Ref.~\cite{obs5}, but now the phenomenology is richer.

\begin{figure*}[t]
\begin{center}
  \includegraphics[type=pdf,ext=.pdf,read=.pdf,width=7.5cm]{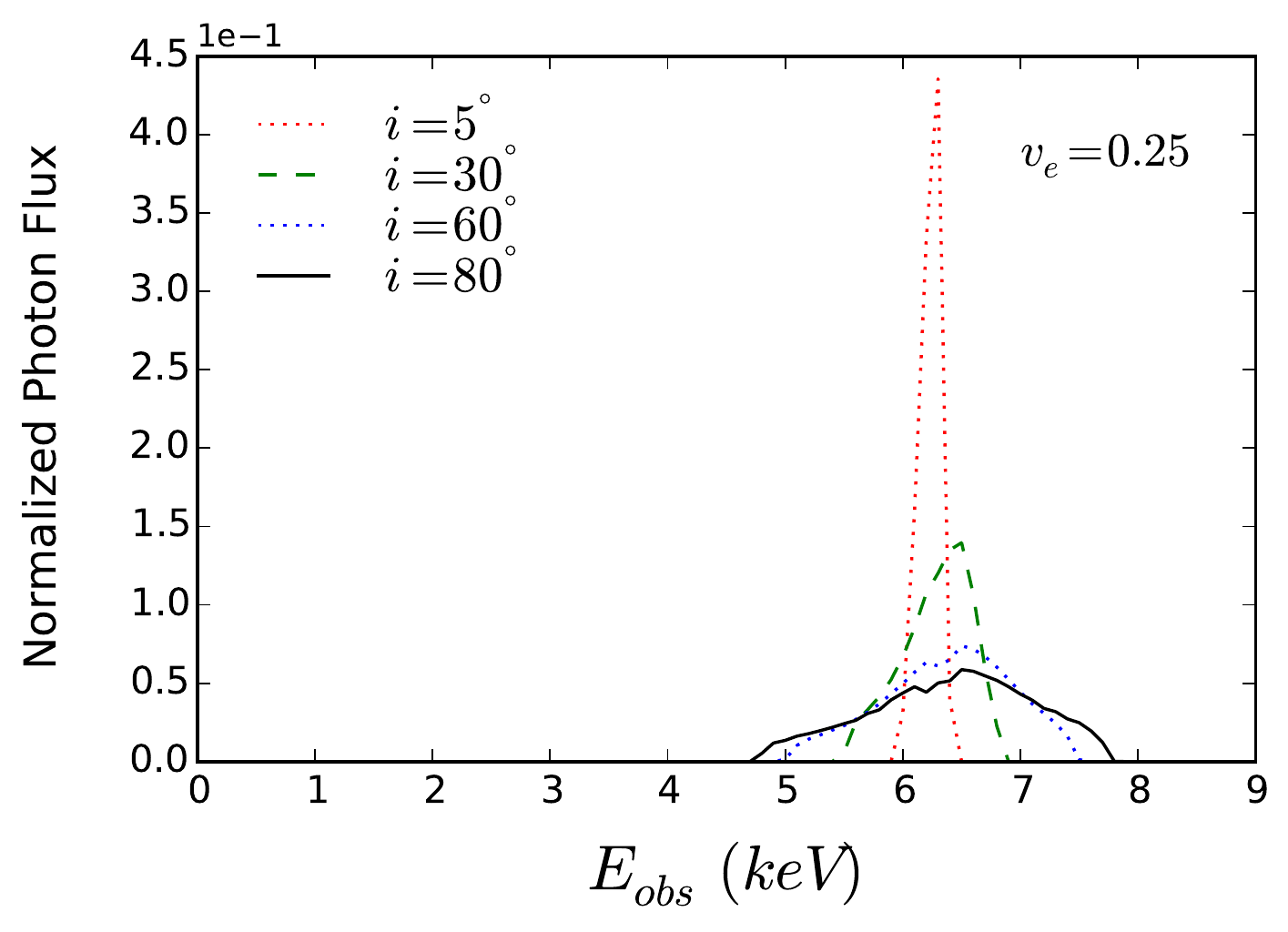}
\hspace{0.8cm}
\includegraphics[type=pdf,ext=.pdf,read=.pdf,width=7.5cm]{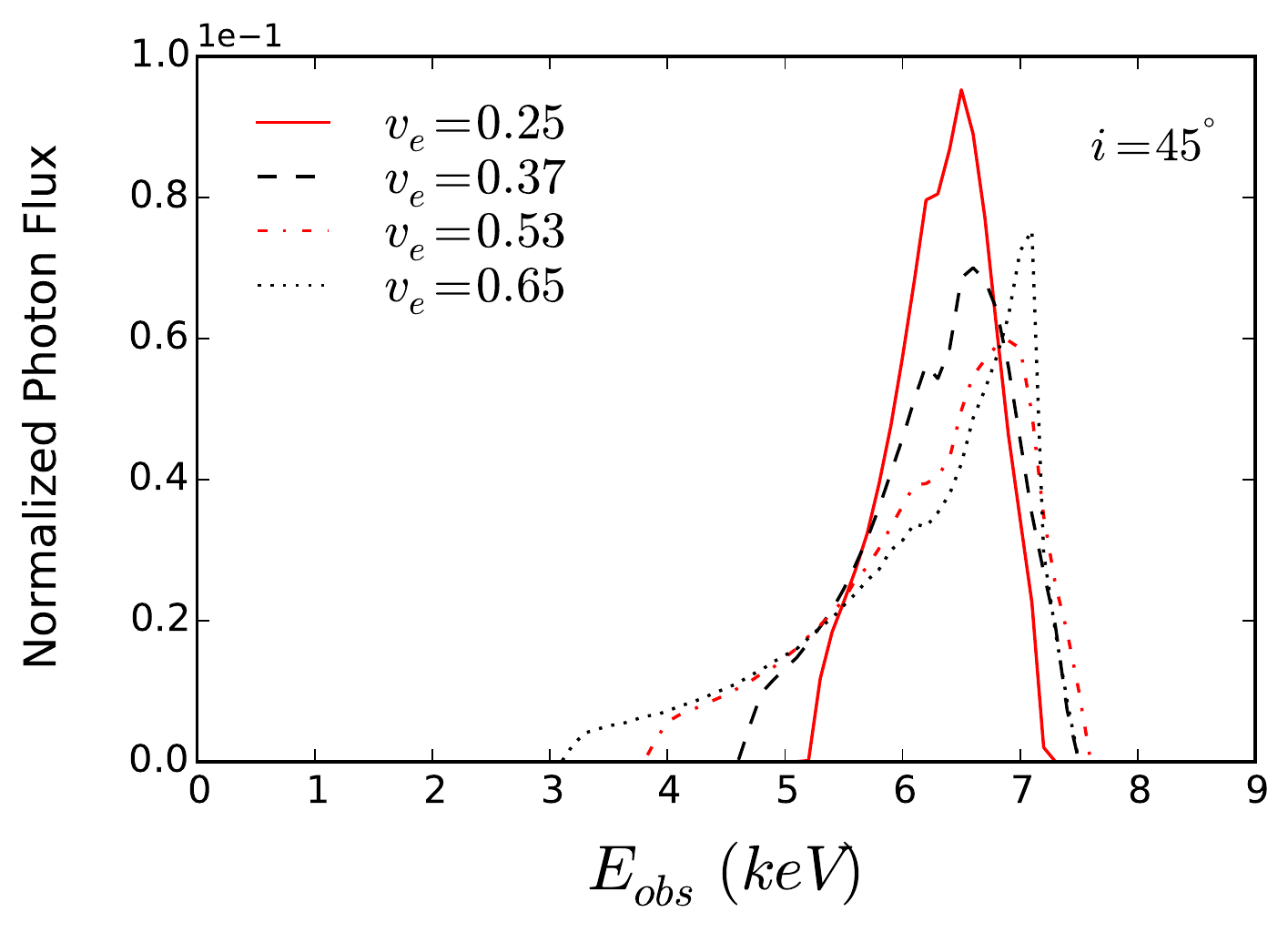}
\hspace{0.8cm}
\includegraphics[type=pdf,ext=.pdf,read=.pdf,width=7.5cm]{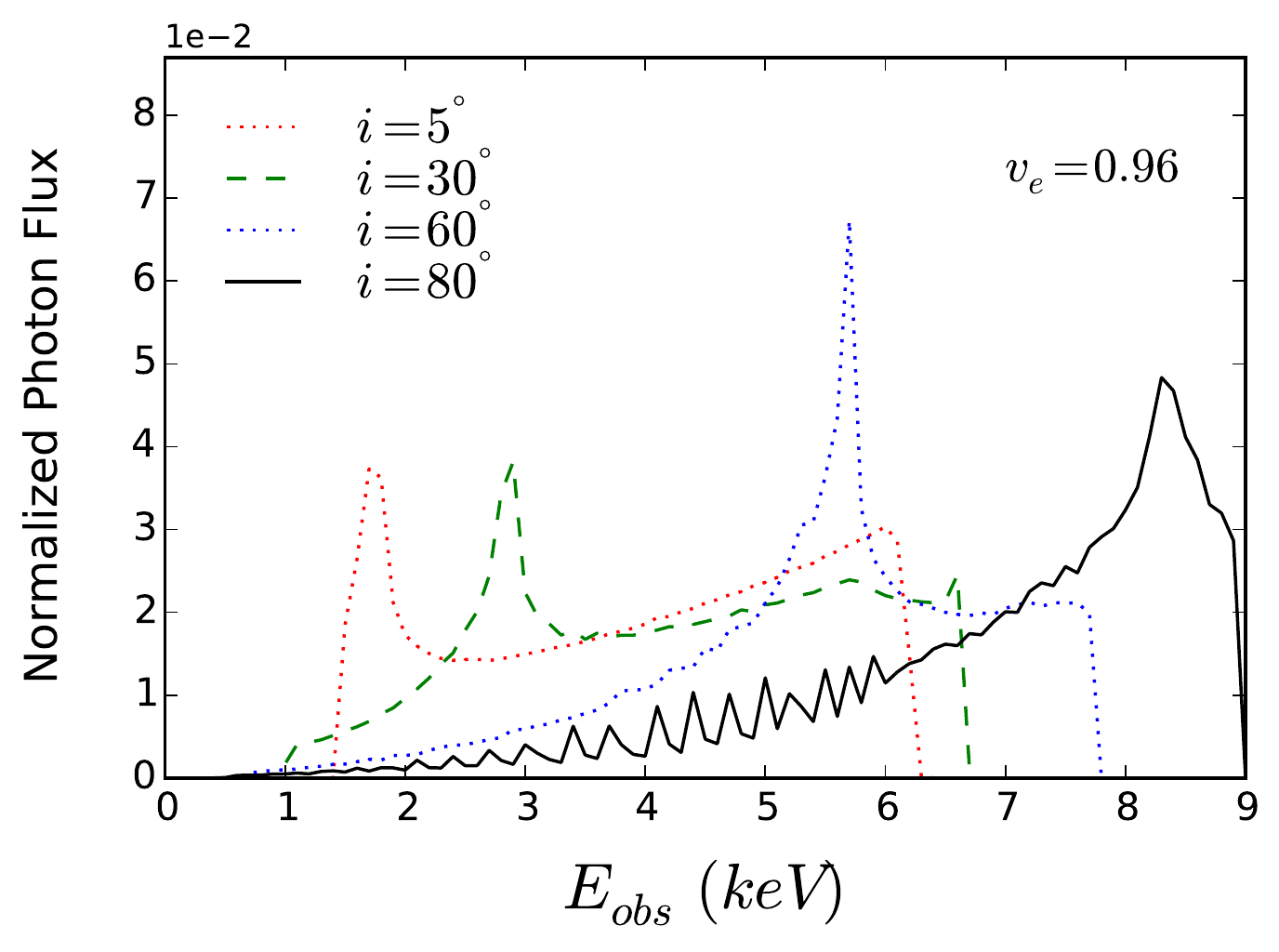}
\hspace{0.8cm}
\includegraphics[type=pdf,ext=.pdf,read=.pdf,width=7.5cm]{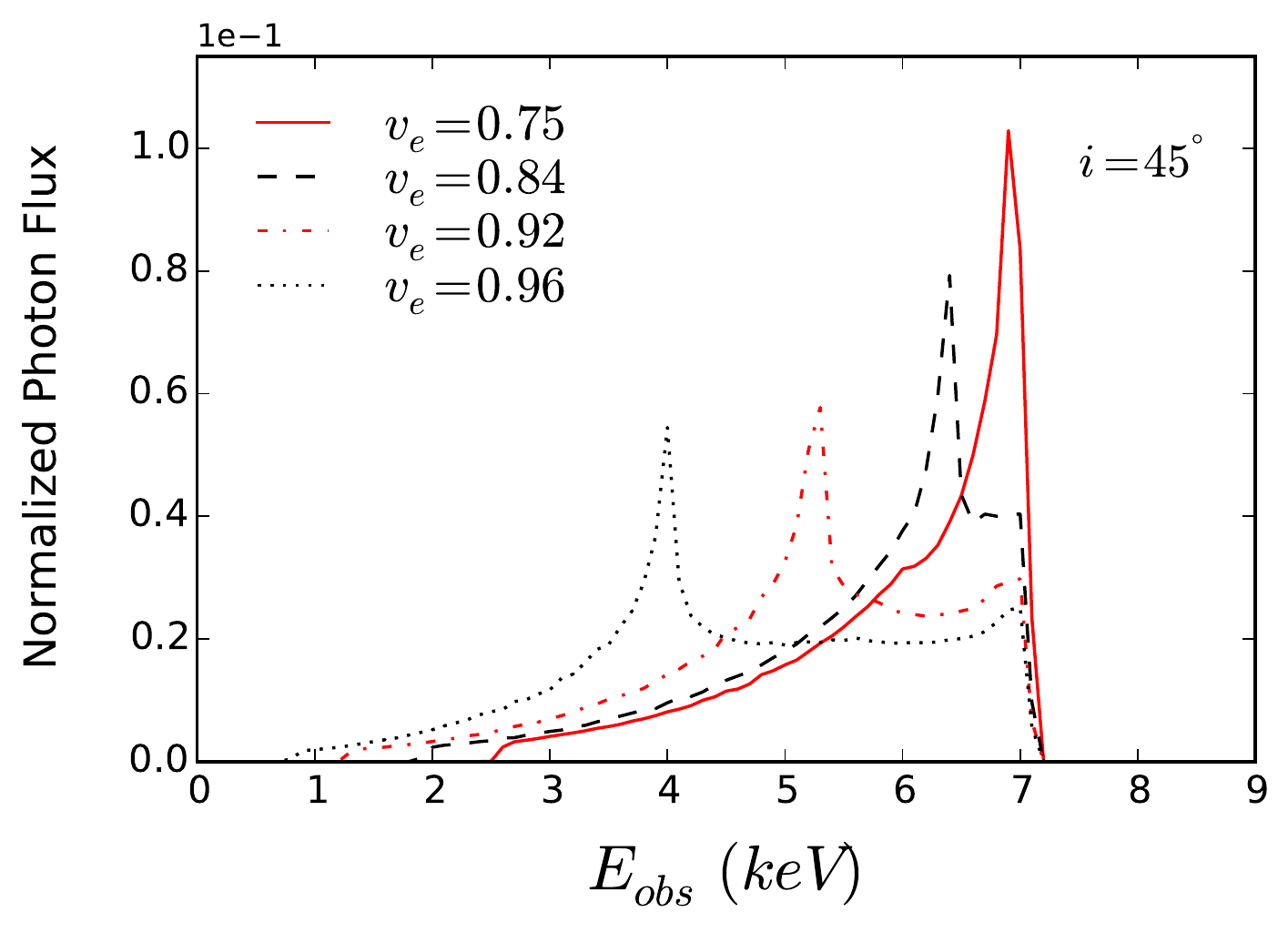}
\end{center}
\vspace{-0.4cm}
\caption{Iron line profile from REWHs with low (top panels) and high (bottom panels) values of the dimensionless rotational velocity of the throat $v_e$. The left panels show the impact on the iron line profile of 
the viewing angle $i$ for a WH with $v_e = 0.25$ and 0.96. In the right panel, $i = 45^\circ$ is fixed and we vary $v_e$. In both 
panels $\eta_0 = 1$. See the text for more details. \label{f-b1}}
\end{figure*}

\section{Simulations} \label{sec:Simu}

The iron line profiles of REWHs have been simulated by using the code described in Refs.~\cite{code1,code2}, with the help of a new subroutine to employ the numerical WH metrics found in Ref.~\cite{kk}.

The iron line profile is described by the photon number density measured by a distant observer, which is given by
\be\label{eq-flux}
N (E_{\rm obs}) &=&  \frac{1}{E_{\rm obs}} \int I_{\rm obs} (E_{\rm obs}) d\tilde{\Omega}
\nonumber\\
&=& \frac{1}{E_{\rm obs}} \int g^3 I_{\rm e} (E_{\rm e}) \, d\tilde{\Omega} \, .
\ee
Here $I_{\rm obs}$ and $E_{\rm obs}$ are, respectively, the specific intensity of the radiation and the photon energy at 
the detection point of the distant observer, $I_{\rm e}$ and $E_{\rm e}$ are the same quantities at the emission point 
in the rest-frame of the gas. $d\tilde{\Omega}$ is the element of the solid angle subtended by the image of the disk in 
the observer's sky. $g = E_{\rm obs}/E_{\rm e}$ is the redshift factor. $I_{\rm obs} = g^3 I_{\rm e}$ follows from 
Liouville's theorem~\cite{mtw}. In what follows, when we only consider the iron line profile, the photon flux~(\ref{eq-flux}) is normalized such that
\be
\int N (E_{\rm obs}) \, dE_{\rm obs} = 1\, .
\ee

The plane of the distant observer is divided into a number of small elements. For each element we integrate the photon trajectory backward in time to the point of emission in the disk. The photon initial conditions are those described in Ref.~\cite{code1}. When these null geodesics cross the equatorial plane, we compute the redshift factor $g$. The 4-velocity of the distant observer is $u^\mu_{\rm obs} = (1,0,0,0)$, the 4-velocity of the gas in the disk is $u^\mu_{\rm e} = u^t_{\rm e} (1 , 0, 0, \Omega)$, where $\Omega$ is the Keplerian velocity given by
\be
\Omega = \frac{- \left( \partial_r g_{t\phi} \right) + \sqrt{\left( \partial_r g_{t\phi} \right)^2 - \left( \partial_r g_{tt} \right) \left( \partial_r g_{\phi\phi} \right)}}{\left( \partial_r g_{\phi\phi} \right)} \, .
\ee
From the normalization condition $g_{\mu\nu} u^\mu_{\rm e} u^\nu_{\rm e}= -1$, we can write the redshift factor $g$
\be
g &=& \frac{E_{\rm obs}}{E_{\rm e}} = \frac{-k_\mu u^\mu_{\rm obs}}{-k_\nu u^\nu_{\rm e}}
\nonumber\\
&=& \frac{\sqrt{-g_{tt} - 2 \Omega g_{t\phi} - \Omega^2 g_{\phi\phi}}}{1 + \lambda \Omega} \, ,
\ee
where $k^\mu$ is the photon 4-momentum and $\lambda = k_\phi/k_t$ is a constant of motion along the photon path, which 
is evaluated at the beginning of the calculation of every photon trajectory.

The disk emission is assumed monochromatic (the rest frame energy is $E_{K\alpha}=6.4$ keV) and isotropic with a power-law radial profile
\be
I_{\rm e} (E_{\rm e}) &\propto& \delta \left(E_{\rm e}-E_{\rm K\alpha}\right) \, 
\left(1+\eta\right)^{-q} \, .
\ee
The parameters of the model are: the rotational velocity of the throat $v_{e}$ (which roughly plays the role of angular 
velocity of the zero angular momentum observer (ZAMO) at the horizon in the Kerr BH case), the inclination angle of the 
disk with respect to the line of sight 
of the distant observer $i$, and the index of the intensity profile $q$. In this exploratory work, we set $q=3$, which 
corresponds to the Newtonian limit in the case of a corona with lamppost geometry. The inner edge of the disk is at the 
ISCO radius, while the outer edge of the disk is set at $\eta_{\rm out} = 100$~$\eta_0$.

\begin{figure*}[t]
\begin{center}
\includegraphics[type=pdf,ext=.pdf,read=.pdf,width=7.5cm]{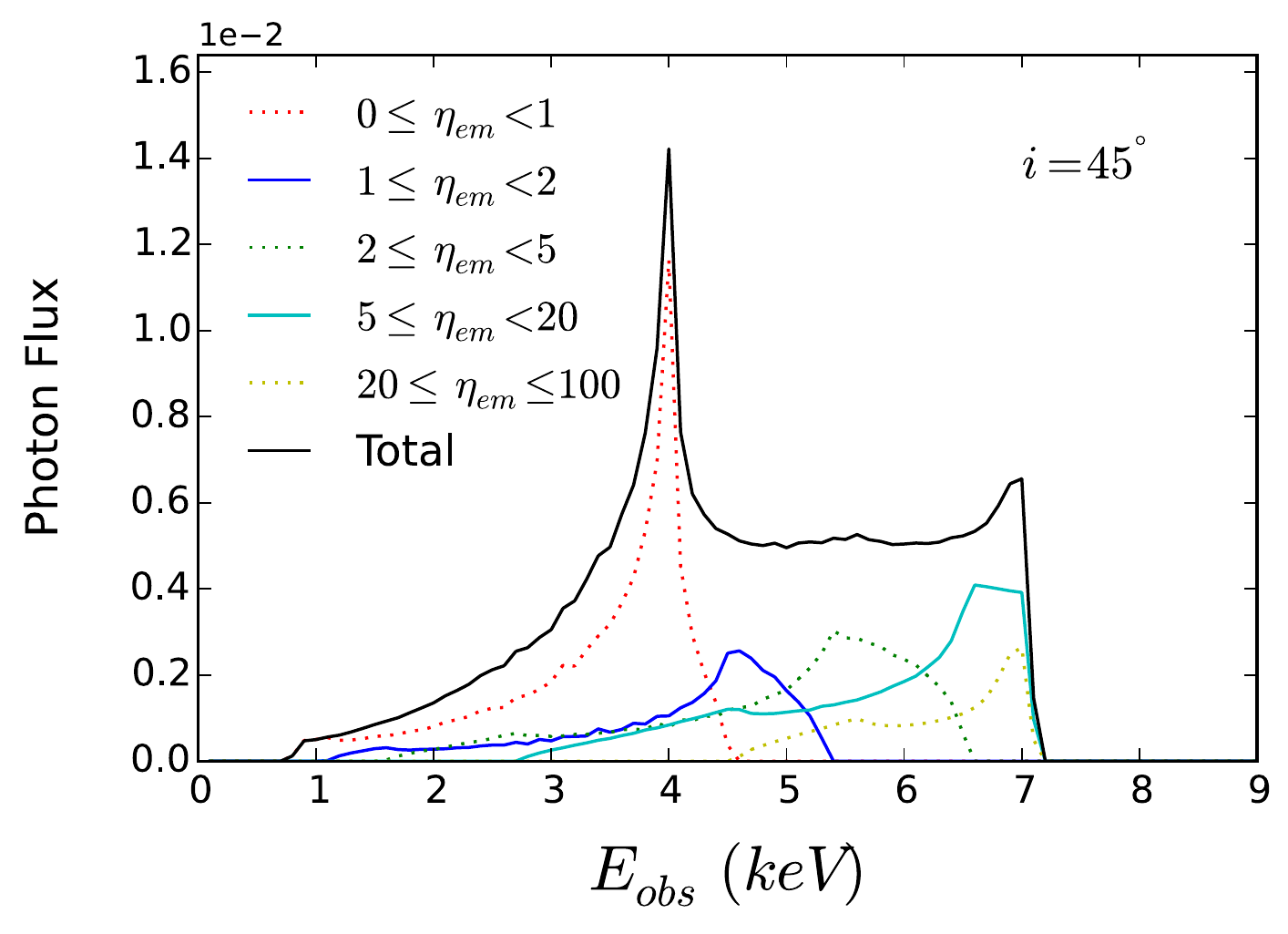}
\hspace{0.8cm}
\includegraphics[type=pdf,ext=.pdf,read=.pdf,width=7.5cm]{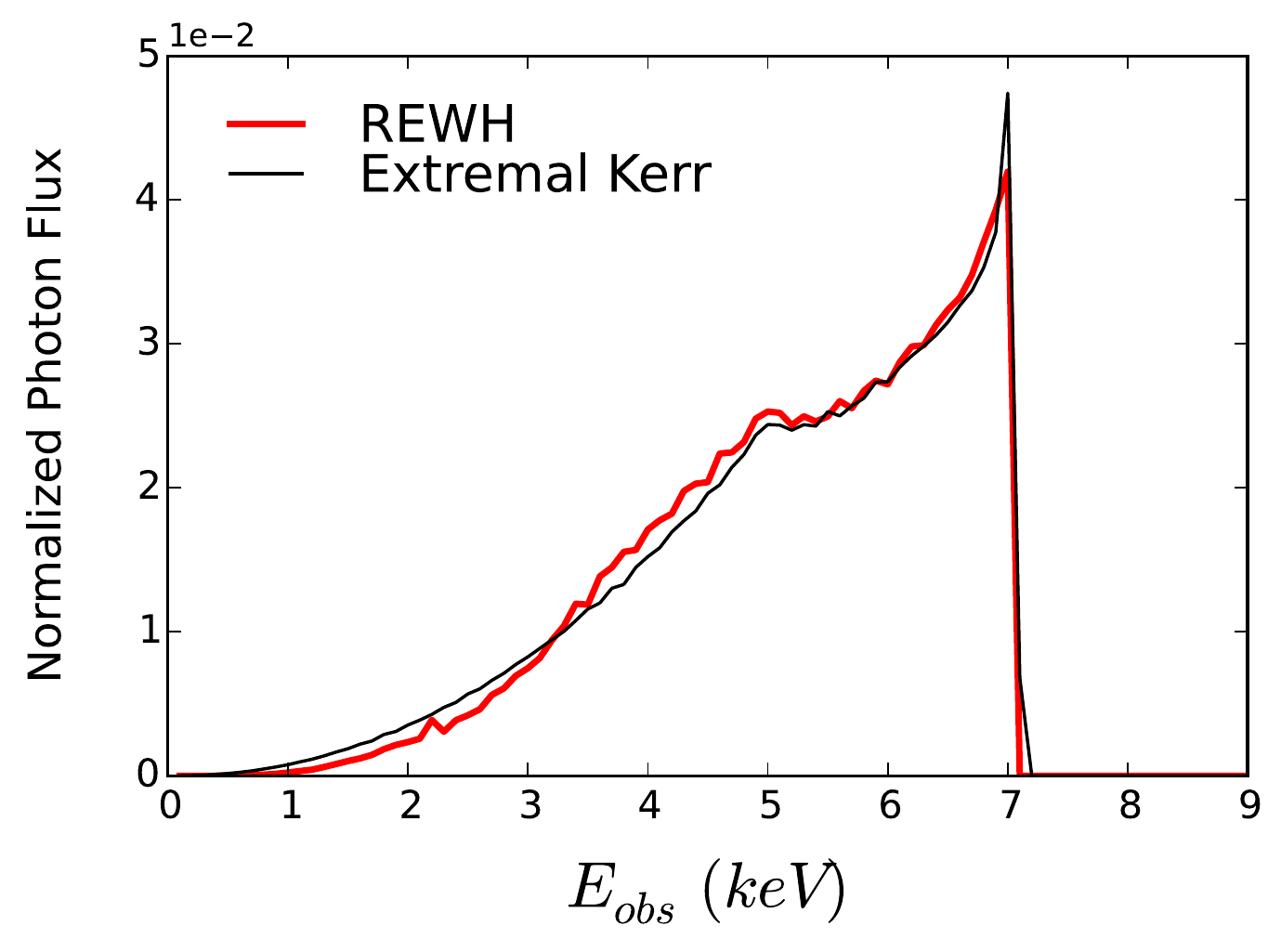}
\end{center}
\vspace{-0.4cm}
\caption{Left panel: contributions to the iron line from radiation emitted from different annuli in the case of a REWH with $v_e = 0.96$, $\eta_0 = 1$, and a viewing angle $i = 45^\circ$. The black solid line is 
the total iron line from an accretion disk with outer radius $\eta_{\rm out} = 100$. The other lines correspond to the 
regions with inner and outer radii reported in the legend. Right panel: comparison between the iron line profile of a 
REWH with $v_e = 0.999$ and $\eta_0 = 0.03$ (thick red curve) and an extremal Kerr BH (thin black 
line). The viewing angle is $i = 45^\circ$. See the text for more details. \label{f-rk}}
\end{figure*}

\section{Discussion} \label{sec:Discu}

\subsection{Iron line profiles}

The iron line profiles in the WH spacetimes are shown in Fig.~\ref{f-b1}. The left panels show 
the impact of the viewing angle $i$ for REWHs with, respectively, $v_e=0.25$ and $v_e=0.96$. In both cases $\eta_0=1$. For $v_e=0.25$, if $i$ is small, we have just a peak 
centered at an energy slightly lower than 6.4~keV, which is the value assumed in the gas rest-frame. As $i$ increases, 
the line becomes broader and develops two peaks, due, respectively, to the Doppler blueshift and redshift. The
interpretation is straightforward. These REWHs are similar to non-rotating Ellis WHs, with the substantial difference 
that they have a mass and bound orbits, so they can have an accretion disk (no accretion disk is possible for 
non-rotating Ellis WHs). However, the gravitational redshift is very weak, so the line is centered at an energy just 
slightly lower than $E_{\rm K\alpha}$. The dominant relativistic effect is the Doppler boosting due to the gas motion, 
which clearly depends on the viewing angle. For the case $v_e=0.96$, we see a prominent peak that moves to higher 
energies as $i$ increases; the interpretation of this line is not straightforward.

While iron line profiles generated from the inner part of the accretion disk of Kerr BHs are not like those in the 
top left panel in Fig.~\ref{f-b1}, this does not automatically mean that it is easy to distinguish a REWH from a Kerr 
BH. Narrow iron lines are common in the X-ray spectrum of BH candidates, but they are usually interpreted as generated 
at larger distances, e.g. from passing clouds, and not belonging to the reflection spectrum. 
Observations of iron line eclipses can distinguish iron lines generated from the inner part of the accretion disk from 
those generated from orbiting clouds at larger distances~\cite{eclipse1,eclipse2}, and therefore in this case they could 
help to test the nature of the compact object. Even the small shift with respect to the 6.4~keV, in the case of AGN may 
be mimicked by the cosmological redshift of the source.

The right panels of Fig.~\ref{f-b1} show the impact of $v_e$ on the iron line profiles of the WHs. 
As $v_e$ increases (with $\eta_0$ fixed), the line develops a low energy tail, which is characteristic of the iron 
line profile in Kerr BHs and is due to the gravitational redshift of the radiation emitted at small radii, where the 
gravitational force is stronger. For low and middle values of $v_e$, the extension of the low energy tail of the WH line is modest, despite the fact that the ISCO radius 
is at the WH throat, which means the gravitational force is much weaker than near a BH. This is completely 
understandable, and indeed there is no horizon in the case of REWHs. In the end, the iron line of a REWH shown in 
Fig.~\ref{f-b1} with moderately high $v_e$ can look like that of a Schwarzschild or slow-rotating Kerr 
BH, in which the low energy tail does not extend to very low energies just because the ISCO radius is not too close to 
the BH horizon.

The shape of the iron line profile of the REWHs with much higher $v_e$, bottom panels of Fig.~\ref{f-b1}, can be 
understood by decomposing the spectrum according to the emission radius in the disk. In the left panel of 
Fig.~\ref{f-rk}, the black solid line represents the total iron line of a REWH with $v_e = 0.96$ and $i = 45^\circ$. The 
other curves represent the spectrum emerging from the five different regions in which we have decomposed the accretion 
disk. The first region of the disk is the annulus with inner edge at the WH throat and outer edge at $\eta = 1$; its 
contribution to the total flux is the red dotted line. The second region is the annulus with inner edge equal to the 
outer edge of the first annulus and outer edge at $\eta = 2$; the radiation of the second region is shown by the blue 
solid line. The third region is the annulus between $\eta = 2$ and $\eta = 5$, the fourth region has $5 < \eta < 20$, 
and the last region has $20 < \eta < 100$, where $\eta = 100$ is the outer edge of the disk. 

\begin{figure*}[t]
\begin{center}
\includegraphics[type=pdf,ext=.pdf,read=.pdf,width=16.0cm]{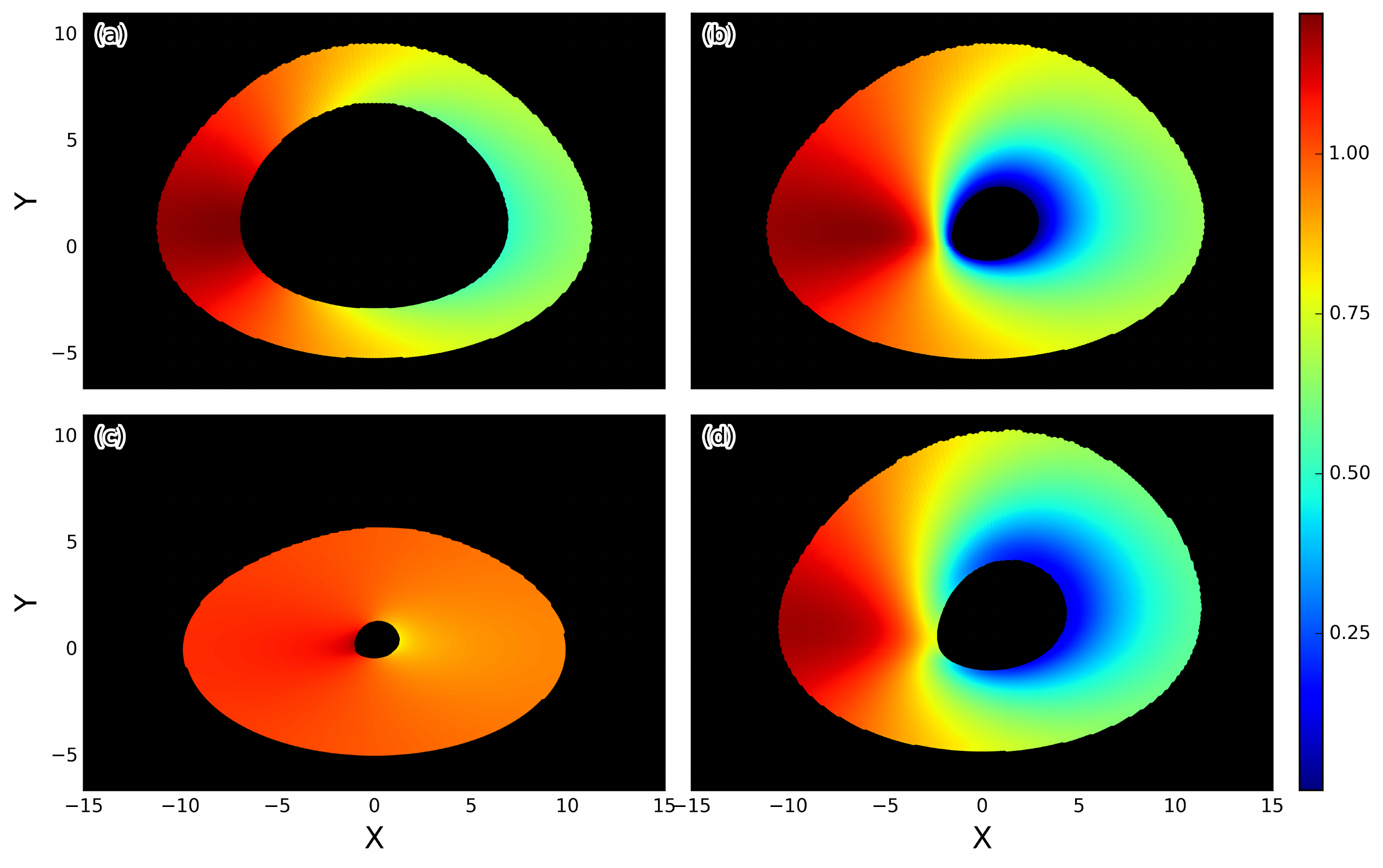}
\end{center}
\vspace{-0.4cm}
\caption{Images of a thin accretion disk around a Schwarzschild BH (a), an extremal Kerr BH (b), a REWH with $v_e 
= 0.25$ and $\eta_0 = 1$ (c), and a REWH with $v_e = 0.96$ and $\eta_0 = 1$ (d). The viewing 
angle is $i = 60^\circ$ and the colors are for the redshift factor $g$. The inner edge of the disk is at the ISCO 
radius 
and the outer edge is at $r = 10$~$M$ for BHs and $\eta_{\rm out} = 10$~$\eta_0$ for WHs. The $X$ and $Y$ axes are in units 
$M=\eta_0=1$. \label{f-image}}
\end{figure*}

The emission from the accretion disk around a BH is quite different. First, the emission at small radii never produces a 
peak. This is because the light bending is different and in the case of a BH many photons emitted near the horizon are 
captured by the BH and those escaping to infinity are more significantly redshifted. This explains the peak in the WH 
spectrum and the absence of a peak in the BH case. Second, the plateau at energies above the peak in the WH iron line 
profile is generated at larger radii and the cut-off of the plateau is determined by the maximum Doppler blueshift. The 
latter occurs at $\eta \approx 15-25$. The peak and the plateau of the WH iron line were already found in the iron lines 
of the class of WHs studied in Ref.~\cite{obs5}. While the WH metrics studied in~\cite{obs5} were not solutions of 
specific theoretical models, the fact that they share these two features with the REWHs of Ref.~\cite{kk} suggests 
that a peak and a plateau may be common signatures of Lorentzian WHs, independent of the underlying theory.

Last, we have compared the iron line of a REWH that approaches an extremal Kerr BH with that of an extremal Kerr 
BH. REWHs reduce to extremal Kerr BHs for $v_e \rightarrow 1$ and $\eta_0 \rightarrow 0$ in a singular 
manner\footnote{Note that for any set of parameters $\eta'_0$, $\omega'_0$ there is a family of solutions with 
parameters $\eta_0 =\Lambda \eta'_0$, $\omega_0 =(1/\Lambda) \omega'_0$, related by the scaling of coordinate $\eta = 
\Lambda \eta'$ and the metric function $\omega =(1/\Lambda) \omega'$. All these solutions possess the same rotational 
velocity of the throat $v_e = v'_e$, see Ref.~\cite{kk} for further details.}. The comparison is shown in the right 
panel of Fig.~\ref{f-rk}. The difference is more due to error in the bilinear interpolation used to interpolate the 
numerical REWH solutions than actual differences between the two metrics in this limit.

\subsection{Images of the accretion disk}

Fig.~\ref{f-image} shows the images of a thin accretion disk around a Schwarzschild BH, an extremal Kerr BH and 
two REWHs with $\eta_0 = 1$ and, respectively, $v_e=0.25$ and $v_e=0.96$. The 
viewing angle is always $i = 60^\circ$. The colors of the accretion disks correspond to the redshift factor $g = E_{\rm 
obs}/E_{\rm e}$. While we do not expect to be able to image thin accretion disks in the near future, these images help 
to figure out the differences between BHs and REWHs.

It is clear that general relativistic effects (gravitational redshift and light bending) are very weak in the REWHs 
with small $v_e$ and that the redshift factor $g$ is mainly determined by the motion of the gas in 
the disk. The central hole is very small because the light bending is very weak and the apparent size of the REWH in 
the plane of the distant observer is close to its geometrical size. The fact that the disk is almost symmetric with 
respect to the axis $Y=0$ is also due to the weak light bending. All these features are consistent with the iron line 
profiles shown in the top left panel in Fig.~\ref{f-b1}.

REWHs with higher $v_e$ are definitively similar to BHs. The most evident difference is that the emission near the 
inner edge of the disk is strongly redshifted in the extremal Kerr BH case (the gravitational redshift is dominant with 
respect to the Doppler boosting), while the effect is milder or absent in the WH case (depending on the viewing angle, 
the Doppler boosting may compensate the gravitational redshift). This is indeed consistent with the bottom left panel 
in Fig.~\ref{f-b1}, where the peak is at very low energies only when the viewing angle is very small and the Doppler
boosting is negligible. The presence of the peak depends instead on the light bending and the number of photons reaching the 
distant observer and it is not shown by the image of the redshift factor.

\begin{table*}
 \centering
\begin{tabular}{|ccccccc|}
\hline
\hspace{0.5cm} &  $v_e$ & \hspace{0.5cm} & $\chi^2_{\rm red,min}$ ($q = 3$) & \hspace{0.5cm} & $\chi^2_{\rm red,min}$ ($q_1$, $q_2$, $r_{\rm b}$ free) & \hspace{0.5cm} \\
\hline
& 0.25 && 0.95691 && 0.94652 &\\
& 0.37 && 1.12678 && 1.07710 &\\
& 0.53 && 1.05557 && 1.05908 &\\
& 0.65 && 1.01038 && 1.03564 &\\
\hline
& 0.75 && 1.02405 && 1.06906 &\\
& 0.84 && 1.00598 && 1.01195 &\\
& 0.92 && {\bf 1.35157} && 1.09106 &\\
& 0.96 && {\bf 1.70910} && {\bf 1.22895} &\\
\hline
\end{tabular}
\vspace{0.3cm}
\caption{Summary of the analysis of the simulations with XIS/Suzaku. The eight WH iron lines are those shown in the right panels of Fig.~\ref{f-b1}, where $v_e$ ranges from 0.25 to 0.96, $\eta_0=1$, and the viewing angle is $i = 45^\circ$. The exposure time of the simulations is 1~Ms. The simulated spectra have been analyzed with a Kerr model. Values with $\chi^2_{\rm red,min} > 1.2$ are highlighted in boldface. See the text for more details. \label{tab}}
\end{table*}

\begin{figure*}[t]
\begin{center}
\includegraphics[type=pdf,ext=.pdf,read=.pdf,width=16.0cm]{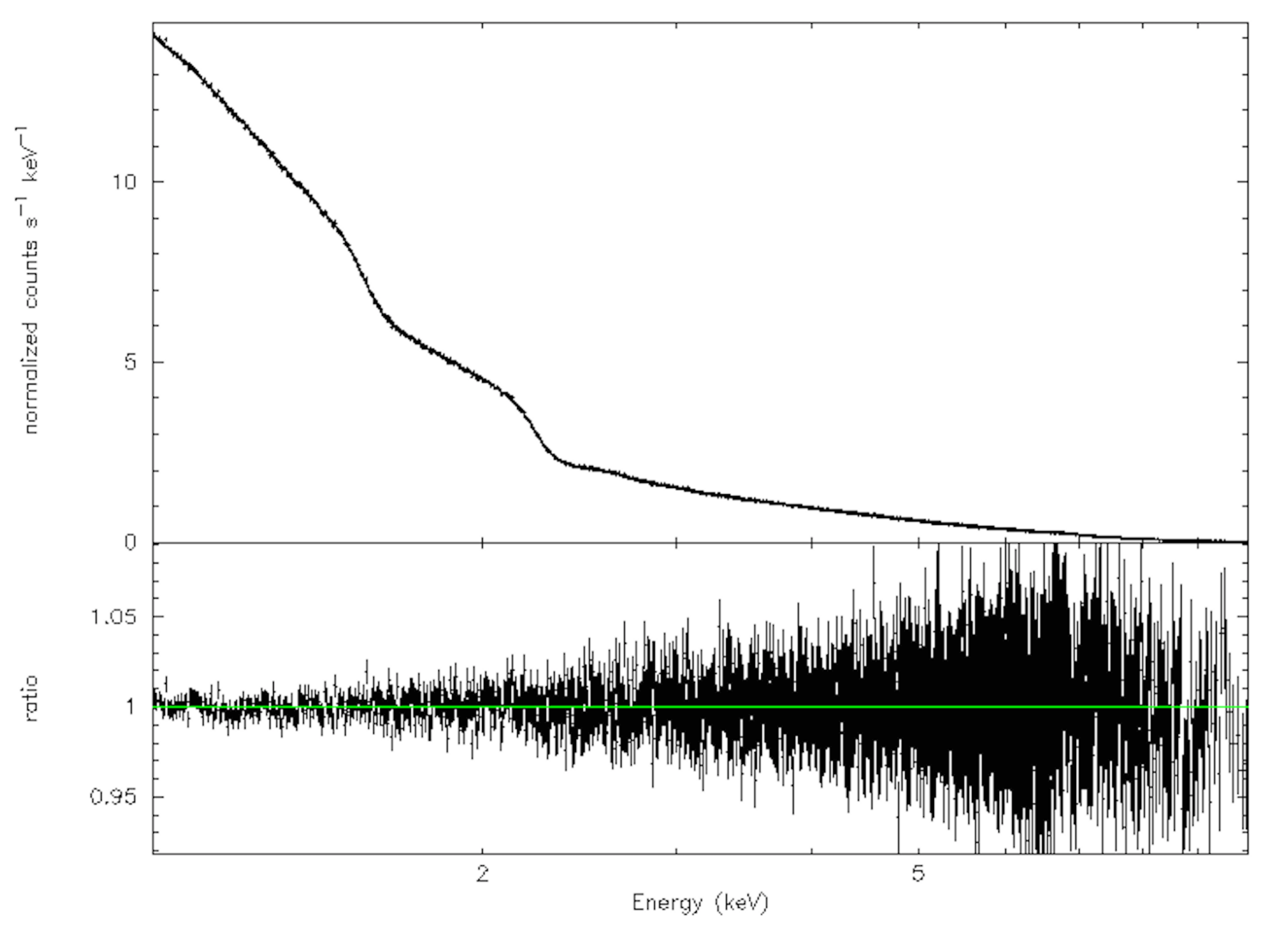}
\end{center}
\vspace{-0.4cm}
\caption{Data and folded spectrum (top panel) and ratio between the data and the Kerr model (bottom panel) for the XIS/Suzaku simulations of the REWH with $v_e = 0.84$. In the Kerr model, $q_1$, $q_2$, and $r_{\rm b}$ are free parameters. $\chi^2_{\rm red,min} = 1.01195$. \label{f-ve84}}
\end{figure*}

\begin{figure*}[t]
\begin{center}
\includegraphics[type=pdf,ext=.pdf,read=.pdf,width=16.0cm]{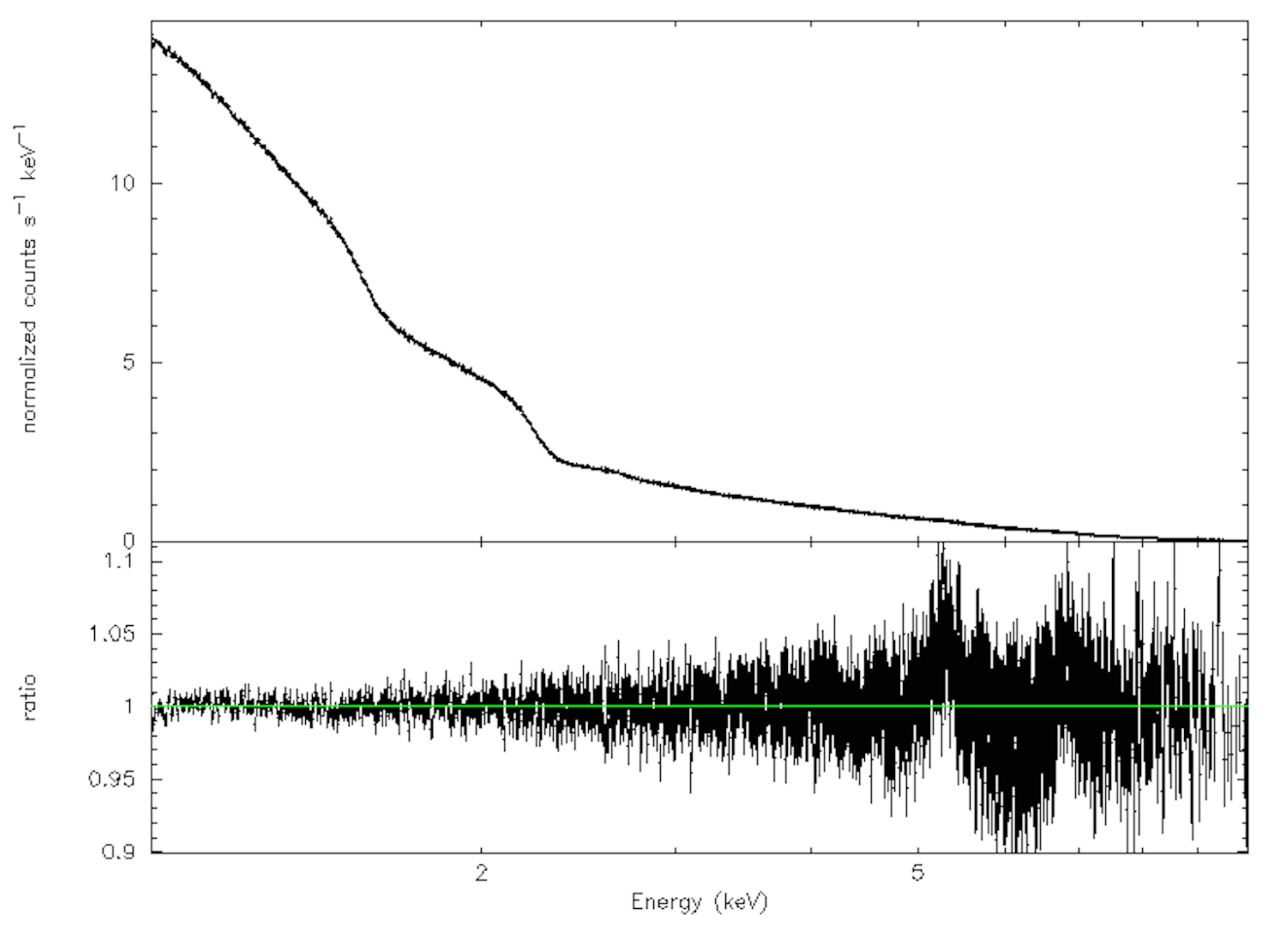}
\end{center}
\vspace{-0.4cm}
\caption{As in Fig.~\ref{f-ve84} for the REWH with $v_e = 0.92$. $\chi^2_{\rm red,min} = 1.09106$ and there is clearly an excess of counts around 5~keV. \label{f-ve92}}
\end{figure*}

\begin{figure*}[t]
\begin{center}
\includegraphics[type=pdf,ext=.pdf,read=.pdf,width=16.0cm]{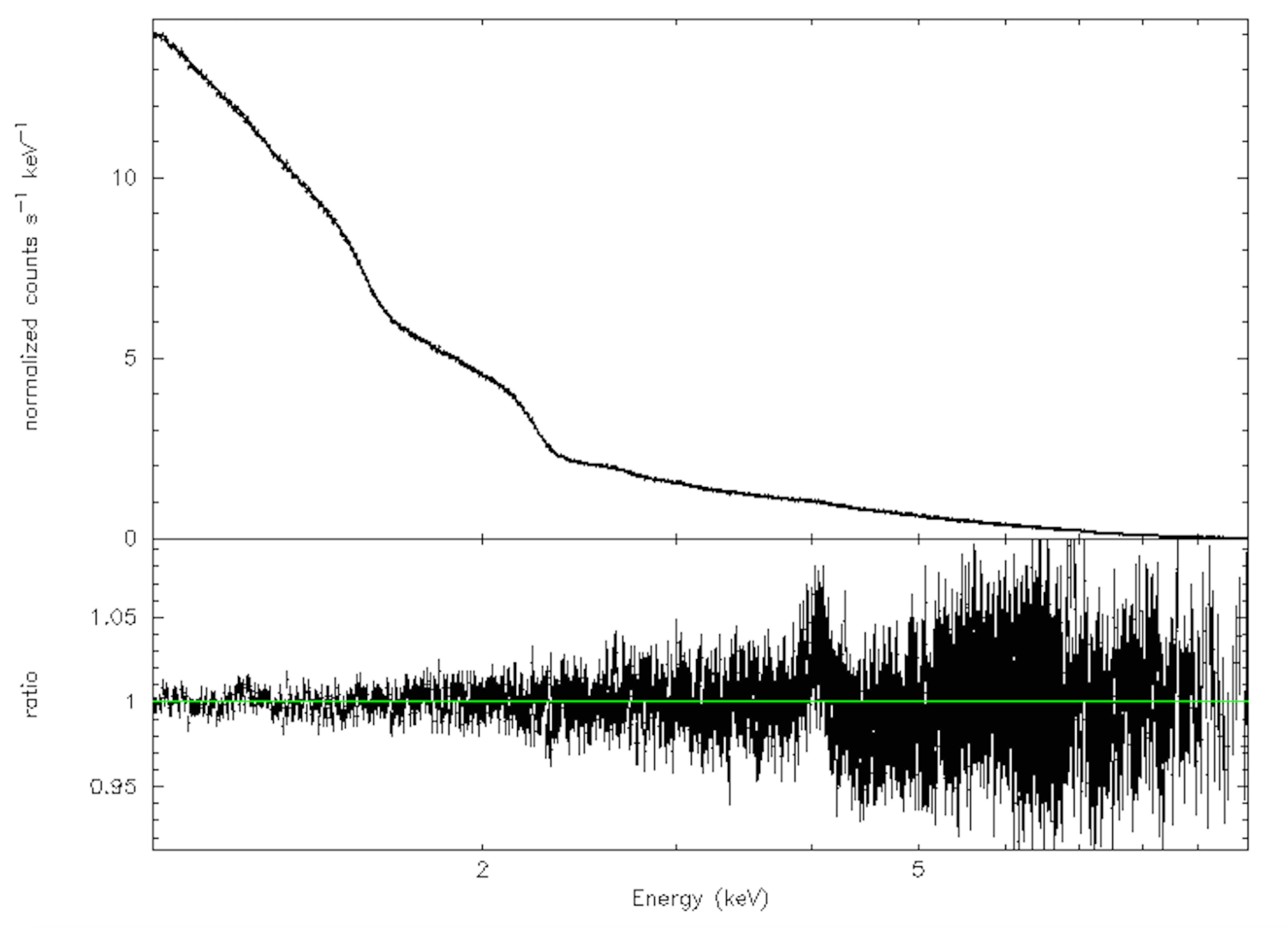}
\end{center}
\vspace{-0.4cm}
\caption{As in Fig.~\ref{f-ve84} for the REWH with $v_e = 0.96$. $\chi^2_{\rm red,min} = 1.22895$ and there is clearly an excess of counts around 4~keV. \label{f-ve96}}
\end{figure*}

\subsection{Simulations with XIS/Suzaku}

In the previous sections, we have studied qualitatively the iron line profile and the disk emission in REWH spacetimes. In this section, we present a simple quantitative analysis to figure out if it is possible to distinguish REWHs and BHs with the available X-ray data. As an explorative study, we simulate some observations with XIS/Suzaku\footnote{http://heasarc.gsfc.nasa.gov/docs/suzaku/} using the WH iron lines computed in the previous sections. The simulated data are then treated as real data and analyzed with a Kerr model. If the fit of the simulated spectrum is good, the interpretation is that the REWH iron line cannot be distinguished from that of a BH. If we cannot find a good fit, current observations should be able to distinguish the two scenarios. Since there are no tensions between the available X-ray data and the theoretical Kerr models, we can argue that the REWHs associated to the spectra with bad fits may be excluded. These results have to be taken with some caution, because we are employing several simplifications. A more detailed analysis with the use of real data and the whole reflection spectrum is well beyond the aim of this preliminary study and is left for the future.

The simulations are done with XSPEC\footnote{http://heasarc.gsfc.nasa.gov/docs/xanadu/xspec/index.html}, using the response file and the background of XIS/Suzaku. We do not consider a particular source, but we simply adopt typical parameters for a bright AGN. The REWH iron line is added to a power law with photon index $\Gamma = 2$. The assumed energy flux in the range 0.7-10~keV is about $2 \cdot 10^{-10}$~erg/s/cm$^2$. The exposure time in our simulations is 1~Ms, which corresponds to a long observation. The resulting total photon count in the range 0.7-10~keV is $\sim 2 \cdot 10^7$. We adopt an iron line equivalent width of about 200~eV. The REWH iron lines are those in the right panels of Fig.~\ref{f-b1}, where $v_e = 0.25$, 0.37, 0.53, 0.65, 0.75, 0.84, 0.92, and 0.96, $\eta_0 = 1$, and the viewing angle is $i = 45^\circ$. After rebinning to assure a minimum photon count per bin of 20, we use XSPEC to fit the spectrum.

The summary of our analyses is shown in Tab.~\ref{tab}. First, we fix the emissivity index $q$ of the iron line profile to 3, which is the actual value used in the simulations of the REWH iron lines. In this case, the free parameters in the fit are: the photon index of the power law $\Gamma$, the normalization of the power law, the normalization of the iron line, the spin parameter of the Kerr BH (because we are fitting the simulated spectra with Kerr models), and the inclination angle of the disk with respect to the line of sight of the distant observer $i$. The minimum of the reduced $\chi^2$ is shown in the second column in Tab.~\ref{tab}.

In order to improve some fits, we assume a more general form of the emission profile. We employ a broken power law with two emissivity indexes, $q_1$ and $q_2$, valid, respectively, for $r < r_{\rm b}$ and $r > r_{\rm b}$, where $r_{\rm b}$ is the breaking radius. Now there are three fitting parameters to model the emissivity, namely $q_1$, $q_2$, and $r_{\rm b}$. Physically speaking, a high value of $q_1$ increases the contribution at small radii and a Kerr model can better mimic the low energy peak in the REWH iron line with high $v_e$; see the discussion in Ref.~\cite{rev1} on this point. The minimum of the reduced $\chi^2$ is shown in the third column in Tab.~\ref{tab}.

We can use two criteria to evaluate the possibility of distinguishing REWHs from BHs: $i)$ the value of the minimum of the reduced $\chi^2$, and $ii)$ the possible presence of unexplained features. As for the point $i)$, if we used the correct theoretical model to fit the data we should find that the minimum of the reduced $\chi^2$ is close to 1, $\chi^2_{\rm red,min} \approx 1.0$. For example, we can exclude the REWHs that, when fitted with a Kerr model, give $\chi^2_{\rm red,min} > 1.2$ (highlighted in boldface in Tab.~\ref{tab}). However, the value of $\chi^2_{\rm red,min}$ only provides an estimate of the quality of the fit of the whole spectrum. An excess or a deficiency of counts in nearby bins at some specific energy is also an indication that our model is missing some important feature. This can be seen by plotting the ratio between the data and the model of the fit.

Figs.~\ref{f-ve84}, \ref{f-ve92}, and \ref{f-ve96} show the data and the folded spectrum (the photon count of the detector according to, respectively, the simulated data and the best fit of the theoretical model) in the top panels and the ratio between the data and the best fit in the bottom panels for the REWH with $v_e = 0.84$, 0.92, and 0.96, $\eta_0 = 1$, and $i = 45^\circ$. For $v_e = 0.84$, we find a good fit ($\chi^2_{\rm red,min} = 1.01195$) and there are no clear unresolved features. For $v_e = 0.92$, the fit is not too bad ($\chi^2_{\rm red,min} = 1.09106$), but there is clearly an excess of counts around 5~keV. For $v_e = 0.96$, the fit is bad ($\chi^2_{\rm red,min} > 1.2$) and the excess of the photon count is around 4~keV.

\section{Summary and conclusions} \label{sec:SumaConc}

The existence of WHs in the Universe cannot be excluded by first principles or by current observations and they are thus 
potentially viable astrophysical structures. Rotating WHs in 4-dimensional general relativity have been found 
numerically only very recently in~\cite{kk}. These objects have a mass and are compact, and they may look like 
astrophysical BHs. In this paper, we have investigated possible observational signatures to identify the REWHs 
of~\cite{kk}. In particular, we have considered the possibility that similar WHs have a thin accretion disk and we have 
studied if the iron line profile in the X-ray reflected spectrum of the disk carries any signature of the presence of 
the central WH.

For small values of the angular velocity at the WH throat, $v_e$, REWHs present a peculiar iron line profile, i.e., the 
gravitational redshift is very weak and the line is broad due to Doppler boosting. In the case of the REWHs with higher 
$v_e$, the iron line has a characteristic peak, which is produced by photons emitted very close to the REWH throat: a 
similar feature is never present in the iron line of a BH due to the different light bending. At energies above this 
peak, the iron line profile shows a plateau, which is also characteristic and is due to the weak gravitational redshift 
and the strong Doppler boosting. The peak and the plateau are two characteristic features that were already found in the 
class of WH metrics studied in~\cite{obs5}. The peak is actually a feature associated with the absence of a horizon and it 
is not a peculiarity of only WHs without a horizon~\cite{malaf}. 

For small values of $v_e$, the iron line cannot develop a long low energy tail, and therefore it looks like a narrow 
line or at best a broad line of a non-rotating or slow-rotating BH. For high values of $v_e$, WHs can capture the main 
features seen in rotating BHs. In the suitable limit to recover extremal Kerr BHs, their iron line profile can be 
literarily indistinguishable from that of a maximally rotating Kerr BH.

We have simulated observations with XIS/Suzaku to figure out if current X-ray missions have the capability of distinguishing REWHs from BHs from the analysis of the iron line. We have assumed typical parameters of a bright AGN and an exposure time of 1~Ms. We have fitted the simulated data with a Kerr model. In most cases, we can find a good fit, namely current observations cannot distinguish REWHs from BHs. However, for high values of $v_e$ ($\eta_0=1$), the REWH iron line cannot be fitted with a Kerr model. Since current observations of AGN are consistent with the Kerr metric, we can exclude that the supermassive BH candidates in galactic nuclei are REWHs with very high $v_e$, say $v_e > 0.9$ (for $\eta_0=1$).


\begin{acknowledgments}
A.C.-A. wishes to thank the Department of Physics at Fudan University, where part of this work was performed, for their 
hospitality. C.B. was supported by the NSFC (grants 11305038 and U1531117), the Thousand Young Talents Program, and the 
Alexander von Humboldt Foundation. 
\end{acknowledgments}


\end{document}